\documentstyle[prl,aps,multicol,amssymb,epsfig]{revtex}
\newcommand{\ket}[1]{\mbox{$|#1\rangle$}}
\newcommand{\bra}[1]{\mbox{$\langle#1|$}}
\title{Quantum Computing in a Macroscopic Dark Period}
\author{Ben Tregenna,$^{1}$ Almut Beige,$^{1,2}$ and Peter L. Knight$^{1}$}
\address{$^{1}$Optics Section, Blackett Laboratory, Imperial College London, London SW7 2BW, England.}
\address{$^{2}$Max-Planck-Institut f\"ur Quantenoptik, Hans-Kopfermann-Stra{\ss}e 1, 85748 Garching, Germany.\footnote{Present address.}}
\date{\today}

\begin{document}
\maketitle

\draft
\begin{abstract}
\begin{center}
\parbox{14.05cm}
{Decoherence-free subspaces allow for the preparation of coherent and entangled qubits 
for quantum computing. Decoherence can be dramatically reduced, yet dissipation is an 
integral part of the scheme in generating stable qubits and manipulating them via one 
and two bit gate operations.  How this works can be understood by comparing the system 
with a three-level atom exhibiting a macroscopic dark period. In addition, a dynamical 
explanation is given for a scheme based on atoms inside an {\em optical cavity} in the 
strong coupling regime and we show how spontaneous emission by the atoms can be highly 
suppressed.} 
\end{center}
\end{abstract}
\noindent
\pacs{PACS: 03.67.Lx, 42.50.Lc}

\begin{multicols}{2}
\section{Introduction}
A major development in recent decades was the realisation that computation is a purely physical process \cite{land}. What operations are computationally possible and with what efficiency depends upon the physical system employed to perform the calculation. The field of quantum computing has developed as a consequence of this idea, using quantum systems to store and manipulate information. It has been shown that such computers can enable an exponential speed up in the time taken to compute solutions to certain problems over that taken by a purely classical device \cite{shor,deutsch,grover}. 

To obtain a quantum mechanical bit (qubit), two well-defined, orthogonal states, denoted by \ket{0} and \ket{1}, are needed. There are certain minimum requirements for any realisation of a universal quantum computer \cite{Vinc}. It must be possible to generate any arbitrary entangled superposition of the qubits. As shown by Barenco {\em et al.}~\cite{barenco}, to achieve this, it suffices to be able to perform a set of universal quantum logic gates. The set considered in this paper consists of the single-qubit rotation and the CNOT gate between two qubits. In addition, the system should be scalable with well characterised qubits and it has to be possible to read out the result of a computation. Finally, the error rates of the individual gate operations should be less than $10^{-4}$ to assure that the quantum computer works fault-tolerantly \cite{PresShor}.

To achieve the required precision, the relevant decoherence times of the system have to be 
much longer than that of a single gate operation and it is this that constitutes the main 
obstacle for quantum computing to overcome. To avoid decoherence it has been proposed that 
decoherence-free (DF) states should be used as qubits. The existence of decoherence-free 
subspaces (DFS) has been discussed widely in the literature by several authors 
(see \cite{Palma,Zanardi97,Lidar98,Guo98,Beigenjp} and references therein). These 
subspaces arise if a system possesses states which do not interact with the environment. 
In addition, the system's own time evolution must not drive the states out of the DFS. 
Recently, the existence of DFSs for photon states has been verified 
experimentally by Kwiat {\em et al.}~\cite{kwiat} and for the states of trapped ions 
by Kielpinski {\em et al.}~\cite{kiel}. 

Far less is known about the manipulation of a system {\em inside} a DFS. One way is 
to use a Hamiltonian which does not excite transitions out of the DFS as has been 
discussed by Bacon {\em et al.}~\cite{manip}.
Alternatively, one can make use of environment-induced measurements \cite{schoen} and 
the quantum Zeno effect \cite{Misra,behe0,zeno} as proposed by 
Beige {\em et al.}~\cite{Beigenjp,usv1} (see also \cite{Dugic}). 
The quantum Zeno effect predicts that any arbitrary but sufficiently weak interaction 
does not move the state of a system out of the DFS, if all non-DF states of the 
system couple strongly to the environment and populating them leads to an immediate 
photon emission. The system then behaves as if it were under continuous observation as 
to whether it is in a DF state or not. Initially in a DF state, the system remains DF
with a probability very close to unity. This idea leads to a realm of new possibilities 
to manipulate DF qubits. 

The possibility of quantum computing {\em using} dissipation has been 
pointed out already by Zurek in 1984 \cite{Zurek} but so far no concrete example for 
a scheme based on this idea has been found. In this paper we discuss in detail such a  
proposal for quantum computing by Beige {\em et al.} outlined in \cite{usv1} and simplify 
its setup. Advantages of this scheme are that it allows for the presence of finite decay 
rates and that its implementation is relatively simple, which should make its experimental 
realisation much less demanding. The precision of gate operations is independent of most 
system parameters and the decoherence times are much longer than the duration of gate 
operations.

In the last few years, many proposals for the implementation of quantum computing have 
been made taking advantage of advances in atom and ion trapping technology. Such methods 
mainly differ in the nature of the coupling between the qubits, e.g.~using collective vibrational 
modes \cite{cp,pcz,molm,schneider}, a strongly coupled single cavity mode 
\cite{domokos,zoller,Zheng,Imamoglu} or 
the dipole-dipole interaction between atoms \cite{brennen,london,briegel}. The physical 
system considered here consists of $N$ atoms (or ions) stored in a linear trap \cite{Lange},
inside an optical lattice \cite{Gomer} or on top of an atomic chip for quantum computing 
\cite{Haensch,Cornell,Schmiedmayer} and interacting via a common cavity
radiation field mode. Each qubit is, as in \cite{pellizzari}, obtained 
from two ground states of an atom, which we call state 0 and 1. The number of qubits is 
thus the same as the number of atoms and the system is scalable.

The single qubit rotations can easily be achieved independently of the cavity by using the well known method of adiabatic population transfer via an excited state \cite{vit,bergmann}. 
Another crucial component of any quantum computer is a mechanism for measuring the qubits in the computational basis. For this a proposal by Dehmelt \cite{shelving} can be used. An additional rapidly decaying level is strongly coupled to one of the ground states with a short laser pulse. The presence or absence of scattered photons then gives an accurate measurement of the atomic state. Further details and extensions of this method are given in \cite{Cook,behe}.

\noindent\begin{minipage}{3.38truein}
\begin{center}
\begin{figure}
\epsfig{file=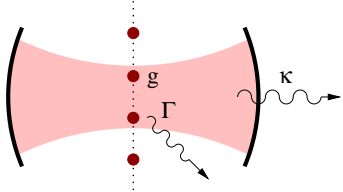, width=4.5cm} \\[0.3cm]
\caption{Schematic view of the atom-cavity system. Two atoms are moved into the cavity where a CNOT gate is performed by the application of a single laser pulse. Here $g$ describes the coupling of each atom to the cavity mode, while $\Gamma$ and $\kappa$ are spontaneous emission rates and cavity damping rate.} \label{fig1}
\end{figure}
\end{center}
\end{minipage}
\vspace*{0.2cm}

To perform a CNOT gate, the two atoms involved have to be moved into a cavity as shown in 
Figure \ref{fig1} and maintained a suitable distance apart to enable laser pulses to address
 each atom individually. The coupling constant of each atom to the cavity mode is denoted 
in the following by $g^{(i)}$. For simplicity we assume here that the coupling strength 
for both atoms is the same and $g^{(1)}=g^{(2)}\equiv g$. To couple non-neighboring atoms, 
ring cavities with a suitable geometry could be used. 

The main source of decoherence in cavity  schemes is the possibility of a photon leaking out of 
the cavity through imperfect mirrors with a rate $\kappa$. Here the qubits are obtained 
from atomic ground states, and so the system is protected against this form of decoherence whilst no gate is performed. Additionally the two atoms in the cavity possess a 
further DF state involving excited atomic levels and an empty cavity. This state is a 
maximally entangled state of 
the atoms and populating it allows the entanglement in the system to change and the CNOT 
gate operation to be realised without populating the cavity mode. To prevent the population of non-DF states, we use the idea described above for the manipulation of a DFS 
which is explained in terms of adiabatic manipulation of DF states. 

The second source of decoherence in the scheme is spontaneous emission from excited atomic states 
which only become populated during a gate operation. The simple 
scheme we discuss in the beginning of this paper involves three-level atoms with a 
$\Lambda$ configuration. It only works with a high success rate if the spontaneous 
decay rate of the upper level is small. More realistically, one can replace 
all the transitions by Raman transitions \cite{vit,bergmann} by 
using three additional levels per atom. We show that the resulting six-level atoms behave 
like the $\Lambda$ systems discussed before but with a highly reduced probability for a 
spontaneous photon emission. As an example we consider $^{40}$Ca$^+$ ions as in the experiment by
Guth\"orlein {\em et al.}~\cite{Lange}.

The paper is organised as follows. In Section II we give a detailed discussion of the 
realisation of the CNOT gate using three-level atoms and show that the behavior of our 
scheme has close parallels with the well-known behavior of a single three-level atom 
exhibiting macroscopic dark periods \cite{shelving}. As will be shown in this paper, 
moving to the correct parameter regime enables the operation to be completed with a high 
success rate and high fidelity of the output. The use of further levels to reduce the 
decoherence from spontaneous emission is covered in Section III. Finally, Section IV, 
offers a summary of our results.

\section{The Realisation of the CNOT gate with a single laser pulse}

To perform a CNOT gate, one has to realise a unitary operation between 
the two qubits involved. This transformation flips the value of the 
target qubit conditional on the control qubit being in state \ket{1}. 
Writing the state of the two qubits as control state followed by target 
state, the corresponding unitary operator equals
\begin{equation}\label{ucnot}
U_{\rm CNOT} = \ket{00}\bra{00}+\ket{01}\bra{01}
+ \ket{10}\bra{11}+\ket{11}\bra{10} ~.
\end{equation}
In this section we discuss a possible realisation of this gate. First an intuitive 
explanation is given, followed by an analytic derivation of the time evolution of the system. 
The success rate of a single gate operation and its fidelity 
under the condition of no photon emission are calculated.

\noindent
\begin{minipage}{3.38truein}
\begin{center}
\begin{figure}
\epsfig{file=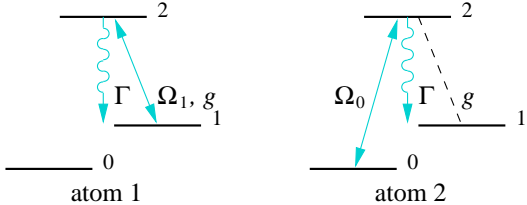, width=7cm} \\[0.4cm]
\caption{Level configuration of the atoms inside the cavity. The ground states 
0 and 1 of each atom form one qubit, while level 2 provides the coupling of
the atoms with coupling constant $g$ with the cavity mode via the 1-2 transition. 
One laser field excites the 1-2 transition of atom 1 with the Rabi frequency 
$\Omega_1$ and another the 0-2 transition of atom 2 with the Rabi frequency 
$\Omega_0$.} \label{lambda}
\end{figure}
\end{center}
\end{minipage}
\vspace*{0.2cm}

To realise a CNOT gate between two qubits the corresponding two atoms are placed 
at fixed positions inside a cavity as shown in Figure \ref{fig1}. To obtain a coupling 
between the atoms via the cavity mode an additional level, level 2, is used.
We assume in the following, that the qubit states $|0\rangle$ and $|1\rangle$  
together with $|2\rangle$ form a $\Lambda$ configuration as shown in Figure \ref{lambda}. 
The 1-2 transition of each $\Lambda$ system couples with the strength $g$ to the 
cavity mode, while the 0-2 transition is strongly detuned. 
In addition two laser fields are required. One laser couples with the Rabi frequency 
$\Omega_1$ to the 1-2 transition of atom 1, the other couples with $\Omega_0$ 
to the 0-2 transition of atom 2 and we choose
\begin{equation} \label{omegacond}
\Omega_{0} =  \Omega_{1} \equiv  \sqrt{2} \Omega ~.
\end{equation}
Note that this choice of the Rabi frequencies is different from the choice in 
\cite{usv1}. There we minimised the error rate, whereas here we are interested in improving
the feasibility of the proposed scheme by simplifying its setup. Only one laser is actually
required per atom! 
 
As in \cite{usv1} we assume in the following that the Rabi frequency 
$\Omega$ is weak compared to the coupling constant $g$ and the decay rate $\kappa$. 
On the other hand, $\Omega$ should not be too small because otherwise spontaneous 
emission from level 2 during the gate operation cannot be neglected. This leads to the 
condition
\begin{eqnarray} \label{cond}
\Gamma \ll \Omega \ll {g^2 \over \kappa} ~~{\rm and}~~ \kappa ~. 
\end{eqnarray}
It is shown in the following that under this condition a laser pulse of duration 
\begin{equation} \label{T}
T = {\sqrt{2}\pi \over \Omega} 
\end{equation}
transforms the initial state of the atoms by a CNOT operation.

In this Section we consider only the two atoms inside the cavity, the laser, the cavity 
field and the surrounding free radiation fields. In the following we denote the energy 
of level $x$ by $\hbar \omega_x$, the energy of a photon with wave number $k$ 
by $\hbar \omega_k$ and $\omega_{\rm cav}$ is the frequency of the cavity field with
\begin{equation}
\omega_{\rm cav} = \omega_2-\omega_1~.
\end{equation}
The annihilation operator for a photon in the cavity mode is given by $b$, and for a photon 
of the free radiation field of the mode $({\bf k},\lambda)$ by $a_{\bf k\lambda}$ or 
$\tilde{a}_{\bf k\lambda}$, for those coupled to the atom or cavity respectively. (The geometry of the setup requires separate fields for each.) 
The coupling of the $j$-2 transition of atom $i$ to the free radiation field can be
described by coupling constant $g_{{\bf k}\lambda}^{(j)}$, whilst $\tilde{g}_{\bf k\lambda}$ 
characterises the coupling of the cavity mode to a different free radiation field.
Using this notation, the interaction Hamiltonian of the system with respect to the 
free Hamiltonian can be written as
\begin{equation}
H_{\rm I}= H_{\rm at-cav}+H_{\rm cav-env}+H_{\rm at-env} + H_{\rm laser\, I}~,
\end{equation}
where 
\begin{eqnarray}\label{hinter}
H_{\rm at-cav}&=& {\rm i} \hbar g \sum_i 
\big[ \ket{2}_i \bra{1} \, b - {\rm h.c.} \big] \nonumber\\
H_{\rm cav-env} &=& {\rm i} \hbar
\sum_{\bf k\lambda} \tilde{g}_{\bf k\lambda} \,
\big[ {\rm e}^{{\rm i}(\omega_1 - \omega_{\bf k})t} \,
b^{\dagger} \tilde{a}_{\bf k\lambda} - {\rm h.c.} \big] \nonumber\\
H_{\rm at-env}&=& {\rm i} \hbar \sum_{i,j} \sum_{\bf k\lambda} 
g_{{\bf k}\lambda,j} \, \big[ {\rm e}^{{\rm i}(\omega_j - \omega_{\bf k})t}
\, \ket{2}_i \bra{j} \, a_{\bf k\lambda} - {\rm h.c.} \big] \nonumber \\
H_{\rm laser \, I} &=& {\textstyle{1 \over 2}} \sqrt{2} \, \hbar \Omega \,
\big[ \ket{0}_2\bra{2} + \ket{1}_1\bra{2} + {\rm h.c.} \big] ~.
\end{eqnarray}
These terms describe the interaction of the atoms with the cavity mode,
the coupling of the cavity or the atoms, respectively, to the external fields
and the effect of the laser on the atomic state.

\subsection{Quantum Computing in a dark period} \label{A}

In this subsection we provide a simple description of the physical mechanism 
underlying our proposal. To do so we point out that there is a close analogy 
between this scheme and the single three-level atom shown in Figure \ref{analogy1}(a).
The atom has a metastable level $A$ which is weakly coupled via a driving laser with 
Rabi frequency $\Omega_{\rm w}$ to level $B$. Level $B$ in turn is strongly coupled 
to a rapidly decaying third level $C$. We denote the Rabi frequency 
of this driving $\Omega_{\rm s}$, the decay rate of the upper level $\Gamma_{\rm s}$ 
and assume in the following
\begin{equation} \label{ccc}
\Omega_{\rm w} \ll {\Omega_{\rm s}^2 \over \Gamma_{\rm s}} 
~~{\rm and}~~ \Gamma_{\rm s} ~.
\end{equation}

Let us assume that the atom is initially in the metastable state $|A \rangle$. In the
absence of the strong driving $(\Omega_{\rm s}=0)$ the atom goes over 
into the state $|B \rangle$ within a time $\pi/\Omega_{\rm w}$. If the strong laser 
pulse is applied, the atoms remain in \ket{A} much longer on average, namely
about the mean time before the first photon emission from level $C$, which equals \cite{dark} 
\begin{equation} \label{tdark}
T_{\rm dark} = {\Omega_{\rm s}^2 \over \Omega_{\rm w}^2\Gamma_{\rm s}}
\gg {\pi \over \Omega_{\rm w}} ~. 
\end{equation}
The transition from level $A$ to level $B$ is strongly inhibited, an effect 
known in the literature as ``electron shelving'' \cite{shelving}.
It is also known as a macroscopic dark period and state $|A \rangle$ as
a dark state \cite{dark}. 

\noindent \begin{minipage}{3.38truein}
\begin{figure}
\begin{center}
\epsfig{file=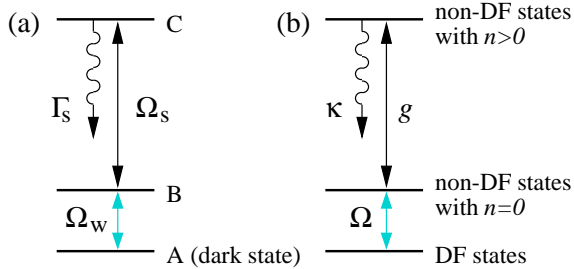, height=3.5cm} \\[0.4cm]
\end{center}
\caption{Analogy between two systems with a macroscopic dark period.
(a) Level scheme of a three-level atom with dark state $|A\rangle$. Here $\Omega_{\rm s}$
is the Rabi frequency of the strong laser driving the 
$B$-$C$ transition, $\Omega_{\rm w}$ is the Rabi frequency of the weak laser driving the $A$-$B$ 
transition and $\Gamma_{\rm s}$ is the decay rate of level $C$.
(b) Schematic view of the level scheme of the two atoms inside
the cavity. The dark state $|A\rangle$ is replaced by the DFS, 
$|B\rangle$ by the subspace of the non-DF states with no photons in the 
cavity $(n=0)$ and $|C\rangle$ by a subspace containing non-DF states with the cavity 
mode populated $(n>0)$. 
The analog to $\Omega_{\rm w}$ is $\Omega$, the analog of $\Omega_{\rm s}$ is 
$g$, and $\Gamma_{\rm s}$ is replaced by $\kappa$.}
\label{analogy1}
\end{figure}
\end{minipage}
\vspace*{0.2cm}

In the scheme we discuss in this paper the levels $A$, $B$ and $C$ are replaced by 
subspaces of states. To show this let us first consider which states play the role 
of the dark state $A$. There are two conditions for dark states or decoherence-free 
(DF) states of a system \cite{Palma,Zanardi97,Lidar98,Guo98,Beigenjp}.
First, the state of the system must be decoupled from the environment. Let us in 
the following neglect spontaneous emission by the atoms inside the cavity by setting 
$\Gamma=0$. Then, this is the case for all states with $n=0$ photons inside the 
cavity. Secondly, the atomic state must be unable to excite the cavity, requiring 
that $H_{\rm at-cav}$ (\ref{hinter}) must annihilate it. The dark states of 
the system are therefore of the form $|\psi\rangle = |0\rangle_{\rm cav} 
\otimes |\varphi\rangle$, where $|\varphi \rangle$ can be an arbitrary superposition 
of the five atomic states \ket{00}, \ket{01}, \ket{10}, \ket{11} and 
the antisymmetric state 
\begin{equation} \label{anti}
|a \rangle \equiv \big[\ket{12}-\ket{21} \big]/\sqrt{2} ~. 
\end{equation}
Here $|n \rangle_{\rm cav}$ denotes the state with $n$ photons inside the 
cavity. The DFS of the two atoms inside the cavity is thus the span of the
individual dark states shown above, resulting in a five-dimensional DFS.

The analog to the shelving system's level $B$ are non-DF states with no photon inside 
the cavity. They are coupled to the DFS via the weak driving laser with Rabi 
frequency $\Omega$. The analog to level $C$ are non-DF states with at least one 
photon in the cavity field. They become excited via coupling of the atoms 
to the cavity mode, with the coupling constant $g$. A photon leaks 
out of the cavity with a rate $\kappa$, which has the same effect as 
the decay rate $\Gamma_{\rm s}$ above. 

Using this analogy, which is summarised in Figure \ref{analogy1}, and replacing condition 
(\ref{ccc}) by condition (\ref{cond}) we can now easily predict the time evolution 
of the two atoms inside the cavity. It suggests that
the weak laser pulse does not move the state of the atoms out of the DFS. 
Nevertheless, the time evolution inside the DFS is not inhibited and is now 
governed by the effective Hamiltonian $H_{\rm eff}$. This Hamiltonian is the projection of 
the laser Hamiltonian $H_{\rm laser \, I}$ with the projector $I\!\!P_{\rm DFS}$
onto the DFS and equals
\begin{equation}
H_{\rm eff} = I\!\!P_{\rm DFS} \,H_{\rm laser \, I}\, I\!\!P_{\rm DFS} ~.
\end{equation}
For the choice of Rabi frequencies made here this leads to the effective 
Hamiltonian
\begin{eqnarray}
H_{\rm eff} &=& {\textstyle{1 \over 2}} \hbar \Omega
\big[ \ket{10}\bra{a}-\ket{a}\bra{11} + {\rm h.c.} \big] 
\otimes |0 \rangle_{\rm cav} \langle 0| ~.
\end{eqnarray}
If the lasers are applied for a duration $T$ as in Eq.~(\ref{T}), then the 
resulting evolution is exactly that desired, the CNOT gate operation. 

The length of the gate operation is chosen such that the additional DF state $|a \rangle$
is no longer populated at the end of the gate operation. It acts as a {\em bus} for 
the population transfer between the qubit states. By populating \ket{a} one can create 
entanglement between the two atoms by applying only a laser field. Note, that 
the cavity always remains empty during the gate operation, nevertheless, it establishes a 
coupling between the qubits.

\subsection{The no photon time evolution}

In this subsection it is shown that the effect of the weak laser fields indeed resembles a 
CNOT operation. We also show that the mean time before the first photon emission
is of the order of $g^2/(\kappa \Omega^2)$ as suggested by Eq.~(\ref{tdark}) and the 
equivalence of the two schemes shown in Figure \ref{analogy1}. To do this we use
the quantum jump approach \cite{HeWi1,HeWi2,HeWi3,HeWi4}. 
It predicts that the state $|\psi^0\rangle$ 
of the two atoms inside the cavity and the cavity field under the condition of 
{\em no} photon emission in $(0,t)$ is governed by the Schr\"odinger equation
\begin{equation}\label{psitcon}
{\rm i} \hbar \, {\rm d}/{\rm d}t \, \ket{{\psi}^0} 
= H_{\rm cond} \, \ket{\psi^0}
\end{equation}
with the conditional Hamiltonian $H_{\rm cond}$. This Hamiltonian is non-Hermitian
and the norm of the state vector \ket{\psi^0} is decreasing in time. From this 
decrease one can calculate the probability for no photon in the time period $(0,t)$, which
is given by
\begin{equation}\label{p0}
P_0 (t,\psi)=\| \, U_{\rm cond} (t,0) \ket{\psi} \|^2~.
\end{equation}
Here we solve Eq.~(\ref{psitcon}) for the laser pulse of Eq.~(\ref{omegacond})
and the parameter regime (\ref{cond}) with the help of an adiabatic elimination 
of the fast varying parameters.

If a photon is emitted, either by atomic spontaneous emission or by cavity decay, then the atomic 
coherence is lost, the gate operation has failed and the computation has to be repeated.
The probability for no photon emission during a single gate operation, $P_0 (T,\psi)$, 
therefore equals the success rate of the scheme. In order to evaluate the quality 
of a gate operation we define the fidelity $F$ of a single gate operation of length 
$T$ as
\begin{equation}\label{F}
F(T,\psi) = {|\langle \psi | U_{\rm CNOT} 
U_{\rm cond}(T,0) |\psi \rangle |^2 \over P_0(T,\psi) } ~.
\end{equation}
This is the fidelity of the scheme under the condition of no 
photon emission. If no photon detectors are used to discover whether the 
operation has succeeded or not, the fidelity reduces to just the numerator.

The conditional Hamiltonian for the atoms in the cavity can be derived from the 
Hamiltonian $H_{\rm I}$ of Eq.~(\ref{hinter}) using second order perturbation 
theory and the assumption of environment-induced measurements on the free 
radiation field \cite{schoen}. This leads to \cite{remark} 
\begin{eqnarray}\label{hcond}
H_{\rm cond} 
&=& {\rm i} \hbar g \sum_i \big[ \ket{2}_i \bra{1} \, b 
- {\rm h.c.} \big] \nonumber \\
& & + {\textstyle{1\over 2}} \sqrt{2} \, \hbar \Omega \,   
\big[ \ket{0}_2\bra{2} + \ket{1}_1\bra{2} + {\rm h.c.} \big] \nonumber \\
& & - {\textstyle{{\rm i}\over 2}} \hbar \kappa \, b^{\dagger} b 
- {\textstyle{{\rm i}\over 2}} \hbar \Gamma \sum_i \ket{2}_i \bra{2} ~.
\end{eqnarray}
The notation we adopt in describing the states of the system is as follows, 
\ket{nx} denotes a state with $n$ photons in the cavity whilst the state of the two 
atoms is given by  $|x \rangle$. Analogously to Eq.~(\ref{anti}) we define
\begin{equation}
|s \rangle \equiv \big[\ket{12}+\ket{21} \big]/\sqrt{2} ~. 
\end{equation}
Writing the state of the system under the condition of no photon emission as
\begin{equation}
\ket{\psi^0} = \sum_{n,x} c_{nx} \, \ket{nx}
\end{equation}
one finds 
\begin{eqnarray} \label{nineqns}
\dot{c}_{n00} & = & 
- {\textstyle{{\rm i} \over 2}} \sqrt{2} \, \Omega \, c_{n02} 
- {\textstyle{1 \over 2}} n \kappa \, c_{n00} \nonumber\\
\dot{c}_{n01} & = & 
- \sqrt{n} g \, c_{n-1 \, 02} 
- {\textstyle{1 \over 2}} n \kappa \, c_{n01} \nonumber\\
\dot{c}_{n10} & = & 
- \sqrt{n} g \, c_{n-1 \, 20}
-{\textstyle{{\rm i} \over 2}} \Omega \, 
\big[ \sqrt{2} \, c_{n20} + c_{na} + c_{ns} \big] \nonumber \\
&& - {\textstyle{1 \over 2}} n \kappa \, c_{n10} \nonumber\\
\dot{c}_{n11} & = & 
- \sqrt{2n} \, g \, c_{n-1\,s}
- {\textstyle{{\rm i} \over 2}} \Omega \, \big[c_{ns} - c_{na}\big]
- {\textstyle{1 \over 2}} n \kappa \, c_{n11} \nonumber\\
\dot{c}_{n02} & = &  
\sqrt{n+1} \, g \, c_{n+1 \, 01}
- {\textstyle{{\rm i} \over 2}} \sqrt{2} \, \Omega \, c_{n00} \nonumber \\
&& - {\textstyle{1 \over 2}} ( n \kappa + \Gamma) \, c_{n02} \nonumber\\
\dot{c}_{n20} & = &  
\sqrt{n+1} \, g \, c_{n+1 \, 10}
-{\textstyle{{\rm i} \over 2}} \sqrt{2} \, \Omega \big[c_{n10}+c_{n22} 
\big] \nonumber \\
&& - {\textstyle{1 \over 2}} ( n \kappa + \Gamma) \, c_{n 20} 
\nonumber\\
\dot{c}_{na} & = & 
- {\textstyle{{\rm i} \over 2}} \, \Omega \, (c_{n10}-c_{n11}+c_{n22})
- {\textstyle{1 \over 2}} ( n \kappa + \Gamma) \, c_{na} 
\nonumber \\
\dot{c}_{ns} & = & 
\sqrt{2(n+1)} \, g \, c_{n+1 \,11} - \sqrt{2n} \, g \, c_{n-1 \,22}\nonumber\\
&&- {\textstyle{{\rm i} \over 2}} \, \Omega \, \big[c_{n10}+c_{n11}+c_{n22} \big]
- {\textstyle{1 \over 2}} ( n \kappa + \Gamma) \, c_{ns} 
\nonumber \\
\dot{c}_{n22} & = & 
\sqrt{2(n+1)} \, g \, c_{n+1\,s}
- {\textstyle{{\rm i} \over 2}} \Omega \,
\big[ \sqrt{2} \, c_{n20} + c_{na} + c_{ns} \big] \nonumber \\
&& - {\textstyle{1 \over 2}} ( n \kappa + 2 \, \Gamma) \, c_{n22} ~.
\end{eqnarray}

There are two different time 
scales in the time evolution of these coefficients, one proportional to $1/\Omega$ 
and $1/\Gamma$ and a much shorter one proportional to $\kappa/g^2$ and $1/\kappa$. 
The only coefficients that change slowly in time are the amplitudes of the DF 
states. All other coefficients change much faster and adapt immediately to the 
system. By setting their derivatives equal to zero we can generate a closed system 
of differential equations for the coefficients of the DF states. 
Neglecting all terms much smaller than $\Omega \kappa/g^2$, 
$\Omega/\kappa$ and $\Gamma/\Omega$ one finds 
\begin{eqnarray} \label{Bdgl}
\left(\begin{array}{c}\dot{c}_{010}\\ \dot{c}_{011}\\  \dot{c}_{0a} 
\end{array}\right) &=& - {\textstyle{1 \over 2}}
\left(\begin{array}{rrr}
10 k_1  &  2 k_1  & {\rm i} \Omega\\
2 k_1   &  2 k_1  & - {\rm i} \Omega\\
{\rm i} \Omega &  -{\rm i} \Omega &  2 k_2 \\ \end{array}\right)
\left(\begin{array}{c} c_{010}\\ c_{011}\\ c_{0a} \\ \end{array}\right) 
\end{eqnarray}
and
\begin{equation}\label{twoeqns}
\dot{c}_{000} = -4 k_1\, c_{000} ~,~ \dot{c}_{001} = 0 
\end{equation}
with
\begin{equation}
k_1 \equiv {\Omega^2\kappa \over 16 g^2} ~,~ 
k_2 \equiv {\Omega^2\kappa \over 16 g^2} + {\Omega^2 \over 2\kappa} + {\Gamma \over 2} ~.
\end{equation}
As a consequence of condition (\ref{cond}) and Eq.~(\ref{T}) we have $k_i T \ll 1$.
Therefore, solving the differential equations (\ref{Bdgl}) and (\ref{twoeqns}) in 
first order in $k_1$ and $k_2$ allows one to describe the effect of the laser 
pulse of length $T$ already to a very good approximation.

By doing so one finds that there is a small population in level $a$ at time $T$. 
This might lead to the spontaneous emission of a photon via atomic decay at which point the CNOT 
operation has failed. With a much higher probability the no photon time evolution 
causes the population of state \ket{0a} to vanish within a time $t_a$ of the 
order of $1/\Gamma$. Taking this into account and assuming that at the begin of 
the gate operation only qubit states are populated we find
\begin{eqnarray} \label{ucond}
U_{\rm cond}(T+t_a,0) &=& U_{\rm CNOT} \nonumber \\ 
& & - {\textstyle {1 \over 4}} \big(6 k_1 - k_2 \big) T
\, \big[ |10\rangle \langle 10| + |11 \rangle \langle 11| \big] \nonumber \\
& & - {\textstyle {1 \over 4}} \big( 10 k_1 + k_2 \big) T 
\, \big[ |10\rangle \langle 11| + |11 \rangle \langle 10| \big] \nonumber \\
& & - 4 k_1 T \, |00\rangle \langle 00| ~.
\end{eqnarray}
If one neglects all terms of the order $k_iT$, then one finds that the no photon time 
evolution of the system is indeed a CNOT operation. In contrast to 
the previous subsection, this has now been derived by solving the time evolution of the 
system analytically. 

From Eq.~(\ref{p0}) and (\ref{ucond}) we find that the success rate of the scheme 
$P_0(T,\psi)$ equals in first order 
\begin{eqnarray}\label{pof}
P_0 (T,\psi) & = & 1 
- {\textstyle {1 \over 2}} (10 k_1 + k_2) T \, 
\big[ |c_{010}|^2+|c_{011}|^2 \big] \nonumber \\
&& - {\textstyle {1 \over 2}} (6 k_1 - k_2) T \, 
\big[ c_{010}c_{011}^* + c_{010}^*c_{011} \big] \nonumber \\
& & - 8 k_1 T \, |c_{000}|^2 ~,
\end{eqnarray}
which is close to unity and becomes
arbitrarily close to unity as $\Omega$ and $\Gamma$ go to zero. In this case
the performance of the gate becomes very slow. Nevertheless, this is successful 
because whilst the gate duration increases as $1/\Omega$, Eq.~(\ref{pof})
shows that the mean time for emission of a photon through the cavity walls scales 
as $1/k_1$ and $1/k_2$ which increases as $1/\Omega^2$.

A main advantage of the scheme we propose here is, that if it works, then the 
fidelity of the gate operation does not differ from unity in first order
of $k_i T$. From Eq.~(\ref{F}) and (\ref{ucond}) we find
within the approximations made above
\begin{eqnarray}\label{f}
F (T,\psi) & = & 1 ~.
\end{eqnarray}
It should therefore be possible that with our scheme the precision of $10^{-4}$ 
can be reached which is required for quantum computing to work fault-tolerantly 
\cite{PresShor}.

\subsection{Numerical results}

In this subsection we present results obtained from a numerical integration 
of the differential equations (\ref{nineqns}). Figure \ref{p0simple} shows the
success rate $P_0(T,\psi)$. For the initial qubit state 
$|10\rangle$ the population of the bus state $|0a\rangle$ during the gate 
operation is maximal and spontaneous emission by the atoms 
the least negligible. We shall therefore use this state as the initial state
to which we apply the gate operation. For $\Gamma \ll \Omega_0$, for which 
$P_0(T,\psi)$ has been derived analytically, a very good 
agreement with Eq.~(\ref{pof}) is found. If the spontaneous decay rate $\Gamma$ 
becomes of the order of $\Omega_0$ then the no photon probability decreases sharply. The reason 
is that the duration $T$ of a single CNOT gate is of the order of $1/\Omega_0$ 
and then also of the order of the life time $1/\Gamma$ of the bus state 
$|0a \rangle$.

The fidelity of the gate operation under the condition of no photon emission through either 
decay channel is shown in Figure \ref{Fsimple}. For $\Gamma=0$ and for the chosen parameters 
the fidelity $F$ is in good agreement with Eq.~(\ref{f}). 
Like the success rate, it only differs significantly from unity if the spontaneous 
decay rate $\Gamma$ becomes of the same order of magnitude as $\Omega_0$. 
A method to prevent spontaneous emission by the atoms is discussed in the 
following section. 

\noindent
\begin{minipage}{3.38truein}
\begin{center}
\begin{figure}
\epsfig{file=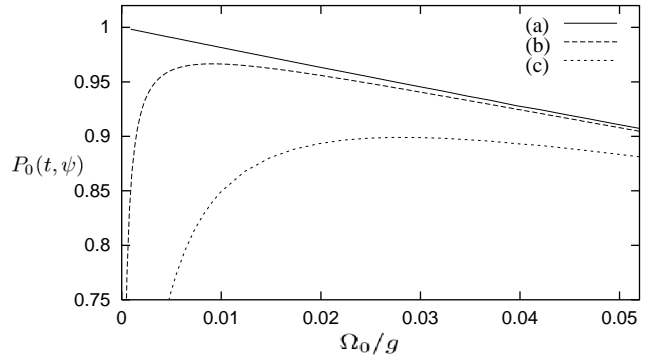, width=3.38truein} \\[0.2cm]
\caption{Success rate of a single gate operation $P_0(T,\psi)$  as a function of the 
Rabi frequency $\Omega_0$ for the initial qubit state $|\psi \rangle = |10\rangle$ and for the  
spontaneous decay rates $\kappa=g$, $\Gamma=0$ (a), $\Gamma=0.0001 \, g$ (b)
and $\Gamma=0.001 \, g$ (c).}
\label{p0simple}
\end{figure}
\end{center}
\end{minipage}
\vspace*{0.2cm}

\noindent
\begin{minipage}{3.38truein}
\begin{center}
\begin{figure}
\epsfig{file=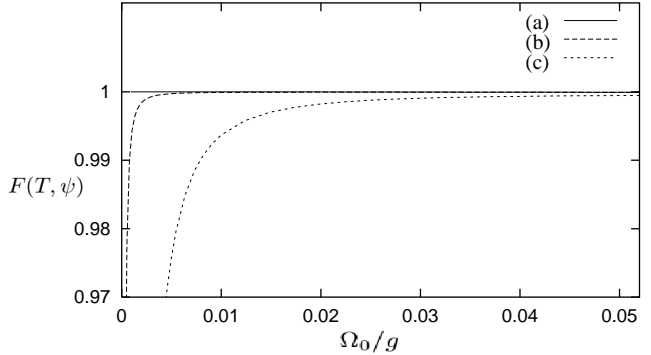, width=3.38truein} \\[0.2cm]
\caption{The fidelity of a single CNOT gate in case of no photon emission 
as a function of the Rabi frequency $\Omega_0$ for 
the same parameters as in Figure \ref{p0simple}.} \label{Fsimple}
\end{figure}
\end{center}
\end{minipage}
\vspace*{0.2cm}

\section{Suppressing spontaneous emission}

The main limiting factor in the scheme discussed in the previous section
is spontaneous emission from level 2. However, we show now how this
can be overcome by replacing all transitions in Figure \ref{lambda} by Raman 
transitions. To be able to do so three additional levels per atom 
are required which we denote in the following by $e_j$. The states $|0\rangle$, 
$|1\rangle$ and $|2 \rangle$ in the new scheme are ground states. They could be obtained, 
for instance, from the $^2$S$_{1/2}$ and $^2$D$_{3/2}$ levels of a trapped calcium ion
as used in Figure \ref{sixy}. 

The new scheme now requires
three strong laser fields applied to both atoms simultaneously, each exciting a 
$j$-$e_j$ transition. Their function is to establish an indirect 
coupling between the states $|0 \rangle$ and $|1 \rangle$ with the state 
$|2 \rangle$ and to generate phase factors. As before, the realisation of a CNOT 
operation requires one transition per atom to be individually addressed. 
One weak laser has to couple only to the $2$-$e_1$ transition in atom 1, and another 
weak one only to the $2$-$e_0$ transition in atom 2. In the following we denote the Rabi 
frequency of the laser with respect to the $i$-$e_j$ transition by $\Omega_{ij}$,
the corresponding detuning of the laser by $\Delta_j$ and the spontaneous 
decay rate of $|e_j \rangle$ is $\Gamma_j$.

\noindent
\begin{minipage}{3.38truein}
\begin{center}
\begin{figure}
\epsfig{file=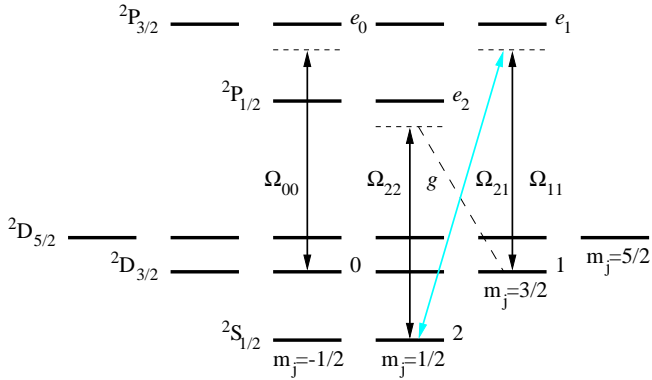, width=3.38truein} \\[0.4cm]
\caption{Level configuration of one of two calcium ions (atom 1) inside the cavity. 
Two of the split $^2D_{3/2}$ levels provide the states 0 and 1 to form one qubit, while level 2 (a $^2S_{1/2}$ state) provides 
the coupling of the atoms via the cavity field. The transition between these 
ground states is realised by Raman transitions via the excited states 
$|e_j \rangle$. The $1$-$e_2$ transition couples to the cavity field with a 
coupling constant $g$ and $\Omega_{ij}$ denotes the Rabi frequency 
of a laser driving the $i$-$e_j$ transition. The configuration of the second atom is similar to the first but with the polarisation of the $\Omega_{21}$ laser reversed so that it now couples states \ket{2} and \ket{e_0}, we call this Rabi frequency $\Omega_{20}$.}
\label{sixy}
\end{figure}
\end{center}
\end{minipage}
\vspace*{0.2cm}

The coupling of both atoms is again realised via the cavity mode which couples to 
the 1-$e_2$ transition of each atom. The frequency of the cavity mode should equal
\begin{equation}
\omega_{\rm cav} = \omega_{e_2} -\omega_1 - \Delta_2 
\end{equation}
such that its detuning is the same as the detuning of the laser driving the 2-$e_2$ 
transition. If desired, the interaction between 
an atom and the cavity can now be effectively switched on or off as required  
by switching on or off the laser which excites the 2-$e_2$ transition, 
relaxing the condition that only the two atoms involved in the CNOT operation can 
be within the cavity. The coupling constant between each atom and the cavity mode is again 
denoted by $g$ and the spontaneous decay rate of a single photon inside the cavity by $\kappa$. 
Using this notation and in the interaction picture with respect to the free Hamiltonian
\begin{eqnarray} \label{hatfree}
H_0 &=&\sum_{i=1}^2 \sum_{j=0}^2 \hbar \omega_j\ket{j}_i\bra{j} 
+ \hbar (\omega_{e_j} - \Delta_j) \ket{e_j}_i\bra{e_j} \nonumber \\
&& + \hbar \omega_{\rm cav} \, b^\dagger b 
+ \sum_{{\bf k} \lambda} \hbar \omega_k 
\big[ \tilde{a}_{\bf k\lambda}^\dagger \tilde{a}_{\bf k\lambda}
+ a_{\bf k\lambda}^\dagger a_{\bf k\lambda} \big] 
\end{eqnarray}
the conditional Hamiltonian becomes
\begin{eqnarray} \label{shcond}
H_{\rm cond} 
&=& {\rm i} \hbar g \sum_i \big[\ket{e_2}_i\bra{1} \, b - {\rm h.c.} \big] \nonumber \\
&& + {\textstyle {1 \over 2}} \hbar \, \big[ \Omega_{21} \ket{2}_1\bra{e_1} 
+ \Omega_{20}\ket{2}_2\bra{e_0} + {\rm h.c.} \big]  \nonumber \\
&& + {\textstyle {1 \over 2}} \hbar \sum_{i,j} 
\big[ \Omega_{jj} \, \ket{j}_i\bra{e_j} + {\rm h.c.} \big] 
+ \hbar \sum_{i,j} \Delta_j \, \ket{e_j}_i\bra{e_j} \nonumber \\
&& - {\textstyle {{\rm i} \over 2}} \hbar \kappa \, b^\dagger b
- {\textstyle {{\rm i} \over 2}} \hbar \sum_{i,j} \Gamma_j \ket{e_j}_i\bra{e_j} ~. 
\end{eqnarray}

\subsection{The no photon time evolution}

In this subsection we determine the parameter regime required for the scheme to behave
as the two atoms in Figure \ref{lambda}
by solving the no photon time evolution of the two six-level atoms inside the cavity. 
It is shown that the difference of the scheme based on six-level atoms compared 
to the scheme discussed in Section II is that the parameters $\Omega_0$, $\Omega_1$ and 
$g$ are now replaced by some effective rates $\Omega_{\rm 0 eff}$, 
$\Omega_{\rm 1 eff}$ and $g_{\rm eff}$ and one has
$\Gamma = 0$. In addition, level shifts are introduced.

First, we should assume that the detunings $\Delta_j$ are much larger than all other system 
parameters. This allows us to eliminate adiabatically the excited states $|e_j\rangle$.
The amplitudes of the wave function of these states change on a very fast time 
scale, proportional $1/\Delta_j$, so that they adapt immediately to the system.
We can therefore set the derivative of their amplitude in the Schr\"odinger equation 
(\ref{psitcon}) equal to zero. Neglecting all terms proportional $1/\Delta_j$ one 
can derive the Hamiltonian $\tilde{H}_{\rm cond}$ which governs the no photon time 
evolution of the remaining slowly varying states. It equals
\begin{eqnarray}\label{shcond2}
\tilde{H}_{\rm cond} 
&=& {\rm i} \hbar \, g_{\rm eff} \sum_i   
\big[ \ket{2}_i \bra{1} \, b - {\rm h.c.} \big] \nonumber \\
&& + {\textstyle{1\over 2}} \hbar 
\big[ \Omega_{\rm 0 eff} \, \ket{0}_2\bra{2} 
+ \Omega_{\rm 1 eff} \, \ket{1}_1\bra{2} + {\rm h.c.} \big] \nonumber \\
&& - {\textstyle{{\rm i}\over 2}}\, \hbar \kappa \, b^{\dagger} b \nonumber \\
&& - \hbar \, {g^2 \over \Delta_2} \sum_i |1\rangle_i \langle 1| \, b^\dagger b
- {\textstyle{1 \over 4}} \hbar \sum_{i,j} {\Omega_{jj}^2 \over \Delta_j} 
\, |j\rangle_i \langle j| \nonumber \\
&& - {\textstyle{1 \over 4}} \hbar \, {\Omega_{20}^2 \over \Delta_0} \, |2\rangle_2 \langle 2|
- {\textstyle{1 \over 4}} \hbar \, {\Omega_{21}^2 \over \Delta_1} \, |2\rangle_1 \langle 2| ~. 
\end{eqnarray}
The first three terms in this conditional Hamiltonian are the same as the terms in 
the Hamiltonian $H_{\rm cond}$ in Eq.~(\ref{hcond}) but with the Rabi frequencies
$\Omega_j$ now replaced by 
\begin{equation} \label{oeff} 
\Omega_{j {\rm eff}} = - {\Omega_{2j} \Omega_{jj} \over 2 \Delta_j} ~,
\end{equation}
the coupling constant $g$ replaced by the effective coupling constant 
\begin{equation} \label{geff}
g_{\rm eff} = - {g \Omega_{22} \over 2 \Delta_2}
\end{equation}
and with $\Gamma=0$. 
The final four terms all represent level shifts. 
The first one of these introduces a level shift to the states $\ket{n1}_i$ with $n>0$, 
while the others correspond to a shift of the states $|0\rangle$,
$|1\rangle$ and $|2\rangle$ of each atom.

To use the setup shown in Figure \ref{sixy} for the realisation of a CNOT gate operation 
we have to assume in analogy to Eq.~(\ref{omegacond}) and (\ref{cond}) that 
\begin{equation} \label{acond}
\Omega_{\rm 0 eff} = \Omega_{\rm 1 eff} 
\end{equation}
and 
\begin{equation} \label{acond2}
|\Omega_{\rm 0 eff}| \ll {g^2_{\rm eff} \over \kappa} ~~{\rm and}~~ \kappa ~.
\end{equation}
By analogy with to Eq.~(\ref{T}) the length $T$ of the weak laser fields with Rabi 
frequency $\Omega_{\rm 0 eff}$ should equal
\begin{equation} 
T = {2\pi \over |\Omega_{\rm 0 eff}|} 
= { 4\pi \Delta_0 \over \Omega_{\rm 20} \Omega_{00}} ~.
\end{equation}
In addition the parameters have to be chosen such that the level shifts 
in Eq.~(\ref{shcond2}) have no effect on the time evolution of the system. If the shifts are appreciable then significant phase differences accrue in the computational basis, leading to a marked decrease in gate fidelity. 
They may be neglected if they are negligible compared to the effective Rabi frequencies 
or the effective coupling constants $g_{\rm eff}$ of the corresponding transition. 
If we choose 
\begin{equation} \label{shifts2}
\Omega_{20} \ll \Omega_{00}~,~~  
\Omega_{21} \ll \Omega_{11} ~~ {\rm and} ~~ g \ll \Omega_{22}~,
\end{equation}
then $g^2/\Delta_2$ becomes negligible compared to $g_{\rm eff}$ and 
$\Omega_{20}^2/\Delta_0$ and $\Omega_{21}^2/\Delta_1$ are much smaller than 
$\Omega_{\rm 0 eff}$ and $\Omega_{\rm 1 eff}$.
For the remaining level shifts we assume that they are the same size for all states.
This is the case if
\begin{equation} \label{shifts}
{\Omega^2_{00} \over \Delta_0} = {\Omega^2_{11} \over \Delta_1}
= {\Omega^2_{22} \over \Delta_2} ~.
\end{equation}
Then they introduce only an overall phase factor to the amplitude of the DF states.

Note, that only the lasers with Rabi frequency $\Omega_{21}$
and $\Omega_{20}$ have to be switched off at the end of a gate operation. 
The setup then resembles that of Section II without 
any laser fields applied and the state of the atoms inside the cavity 
does not change anymore. 

\subsection{Numerical results}

Finally, we present some numerical results for the success rate $P_0(T,\psi)$
for a single CNOT operation and for the fidelity $F(T,\psi)$ to show how well the 
setup shown in Figure \ref{sixy} for the suppression of spontaneous emission in the scheme works. 
The following results are obtained from a numerical integration of the no photon
time evolution with the conditional Hamiltonian $H_{\rm cond}$ given in Eq.~(\ref{shcond}).
For simplicity and as an example we assume in the following 
\begin{equation} \label{scond1}
\Delta_0 = \Delta_1 = \Delta_2 \equiv \Delta
\end{equation}
which implies as a consequence of Eq.~(\ref{shifts}) that
\begin{equation} \label{scond2}
\Omega_{00} = \Omega_{11} = \Omega_{22} \equiv \Omega ~.
\end{equation}
The conditions (\ref{acond}), (\ref{acond2}) and (\ref{shifts2}) given in the previous 
subsection are fulfilled if for instance $\Omega_{20} =\Omega_{21}$, $\kappa = |g_{\rm eff}|$ 
and $\Omega_{20} \ll g \ll \Omega$. In addition, the detuning $\Delta$ should be much larger 
than all other parameters, i.e.~$\Omega \ll \Delta$. For simplicity we assume here that the 
spontaneous decay rates are for all states $|e_j \rangle$ the same, 
\begin{equation} 
\Gamma_0 = \Gamma_1 = \Gamma_2 \equiv \Gamma~.
\end{equation}
The initial state of the qubits is in the following as in Section II given by $|10\rangle$. 

\noindent
\begin{minipage}{3.38truein}
\begin{center}
\begin{figure}
\epsfig{file=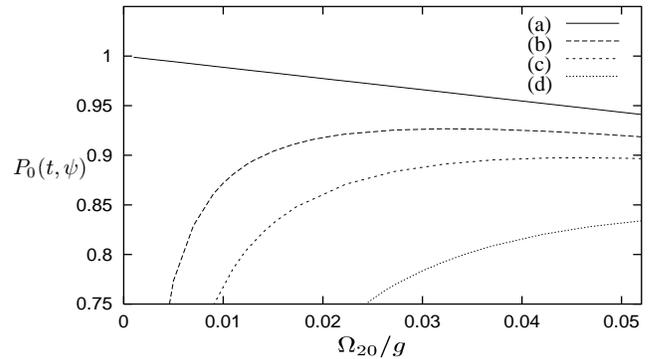, width=3.38truein} \\[0.2cm]
\caption{Probability for no photon emission during a single CNOT operation for the 
initial atomic state $|10\rangle$, different Rabi frequencies $\Omega_{20}$ and 
the spontaneous decay rates $\kappa = |g_{\rm eff}|$, $\Gamma=0$ (a), 
$\Gamma=0.1 \,g$ (b), $\Gamma=0.2 \,g$ (c)  and $\Gamma= 0.5 \, g$ (d). In addition 
is $\Delta = 1000 \, g$ and $\Omega = 2 \,g$.} \label{Phard}
\end{figure}
\end{center}
\end{minipage}
\vspace*{0.2cm}

Figure \ref{Phard} shows the success rate for a single CNOT gate operation. As one can 
see by comparing the results for $\Gamma=0$ to the results for $\Gamma=0$ in 
Figure \ref{p0simple}, the presence of the additional level shifts in Eq.~(\ref{shcond2}) 
increases slightly the no photon probability of the scheme. Otherwise,
it shows the same qualitative dependence on $\Gamma$ 
and $\Omega_0$ or $\Omega_{20}$, respectively, in both Figures.
The main advantage of the scheme using six-level atoms is that the spontaneous emission 
rates of the excited states $|e_j\rangle$ can now be of the same order as the cavity 
coupling constant $g$ without decreasing the success rate of the gate operation 
significantly which allows for the implementation of the scheme with optical cavities.   

The no photon time evolution of the system over a time interval 
$T$ indeed plays the role of a CNOT gate to a very good approximation.
The quality of the gate can be characterised through the fidelity $F$ defined in 
Eq.~(\ref{F}). The fidelity obtained through numerical solution is now very close to unity.
For for the whole range of parameters used in Figure \ref{Phard} it is above $99.8\,\%$.

One could object, that the duration $T$ of the gate presented here 
is much longer than for the gate described in the previous section. 
But, as predicted in Section II.B,
the ratio of the gate operation time to the decoherence time is highly reduced.
One of the main requirements for quantum computers to work fault-tolerantly is for this ratio 
to be low. This is now fulfilled for a much wider range of parameters. 

\section{Conclusions}

We have shown in this paper that it is possible to fulfill all the requirements placed upon a universal quantum computer in a quantum optical regime. We have presented two such schemes, the first is similar to that shown in \cite{usv1} except that it has been optimised for simplicity and its construction is feasible using current experimental techniques. The second suggestion builds on this by substantially reducing the errors arising from spontaneous decay at the expense of slightly increased complexity of implementation. 

By comparing the underlying physical mechanism to that observed in electron shelving experiments, we hope to have shed new light on passive methods of coherence control.

As a first step to test the proposed scheme one could use it to prepare two atoms in 
a maximally entangled state and measure its violation of Bell's inequality 
as described in \cite{bell}.
Finally we want to point out that we think that the idea underlying our scheme can be carried over to other systems and to arbitrary forms of interactions to manipulate their state and so lead to a realm of new possibilities for the realisation of decoherence-free quantum computing.

\vspace*{0.5cm}
{\em Acknowledgment.} 
We thank D. F. V. James, W. Lange, and B.-G. Englert for interesting and 
stimulating discussions. This work was supported by the UK Engineering and 
Physical Sciences Research Council, and the European Union through the projects `QUBITS' 
and `COCOMO'.

\end{multicols}

\begin{references}
\bibitem{remark}
Please note that the definiton of $\kappa$ and $\Gamma$ is different from 
the definition used in \cite{usv1}.
\end{references}
\end{document}